\documentclass[letter,traditabstract]{aa}
\usepackage{epsfig}
\usepackage{times}
\usepackage{natbib}
\bibpunct{(}{)}{;}{a}{}{,}

\newbox\grsign \setbox\grsign=\hbox{$>$} \newdimen\grdimen \grdimen=\ht\grsign
\newbox\simlessbox \newbox\simgreatbox \newbox\simpropbox \newbox\wtildebox 
\setbox\simgreatbox=\hbox{\raise.5ex\hbox{$>$}\llap
     {\lower.5ex\hbox{$\sim$}}}\ht1=\grdimen\dp1=0pt
\setbox\simlessbox=\hbox{\raise.5ex\hbox{$<$}\llap
     {\lower.5ex\hbox{$\sim$}}}\ht2=\grdimen\dp2=0pt
\def\simgreat{\mathrel{\copy\simgreatbox}}
\def\simless{\mathrel{\copy\simlessbox}}

\newcommand{\be}{\mbox{\begin{equation}}}
\newcommand{\ee}{\mbox{\end{equation}}}

\newcommand{\Cref}{\mbox{$m_{\rm ref}$}}

\newcommand{\msun}{\mbox{M$_\odot$}}

\renewcommand{\d}{{\rm d}} %roman d in equations

\begin{document}

\title{Can habitable planets form in clustered environments?
}

\author{M.~de~Juan~Ovelar\inst{1}, J.~M.~D.~Kruijssen\inst{2}, E.~Bressert\inst{3,5}, L.~Testi\inst{3,6}, N.~Bastian\inst{4} and H.~Canovas\inst{7}}

\institute { 
                 {\inst{1}Leiden Observatory, Leiden University,
                 PO Box 9513, 2300RA Leiden, The Netherlands}\\
                 {\inst{2}Max-Planck Institut f\"{u}r Astrophysik,
                 Karl-Schwarzschild-Stra\ss e 1, 85748, Garching, Germany}\\
                 {\inst{3}European Southern Observatory,
                 Karl-Schwarzschild-Stra\ss e 2, 85748, Garching, Germany}\\
                 {\inst{4}Excellence Cluster Universe,
                 Boltzmannstra\ss e 2, 85748 Garching, Germany}\\
                 {\inst{5}School of Physics, University of Exeter, 
                 Stocker Road, Exeter EX4 4QL, UK}\\
                 {\inst{6}INAF-Osservatorio Astrofisico di Arcetri, 
                 Largo E. Fermi 5, I-50125 Firenze, Italy}\\
                 {\inst{7}Departamento de F\'{i}sica y Astronom\'{i}a, Universidad de Valpara\'{i}so,
                 Avda. Gran Breta\~{n}a 1111, Valparaiso, Chile}                 
                 }

\date{Received date 18 May 2012; accepted date}

\abstract{{We present} observational evidence of {environmental effects on the formation and evolution of planetary systems}. We combine catalogues of resolved {protoplanetary} discs (PPDs) and young stellar objects in the solar neighbourhood to analyse the {PPD} size distribution as a function of ambient {stellar} density. By running Kolmogorov-Smirnov tests between the {PPD} radii {at different} densities, we find empirical evidence, at {the $>97\%$ confidence level,} for {a change in the {PPD} radius distribution} at {{ambient stellar densities} $\Sigma\simgreat 10^{3.5}~{\rm pc}^{-2}$. This coincides with a simple theoretical estimate for the truncation of {PPDs} {or planetary systems} by dynamical encounters. If this agreement is causal, the ongoing disruption of {PPDs} and planetary systems limits the possible existence of planets in the habitable zone, with shorter lifetimes at higher host {stellar masses and ambient densities}. {Therefore,} {habitable planets} are not likely to be present in long-lived stellar clusters{, and may have been ejected altogether to form a population of unbound, free-floating planets}. We conclude that, while highly suggestive, {our results} should be verified through other methods. Our simple model shows that truncations should lead to a {measurable} depletion of the {{PPD}} mass function {that can} be detected with ALMA observations of the densest nearby and young clusters.}}

\authorrunning{M.~de~Juan~Ovelar et al.}
\titlerunning{Planets in clustered environments}

\maketitle

\section{Introduction} \label{sec:intro}
{For the past} decade, exoplanetary systems are being discovered at a spectacular rate \citep[e.g.][]{mayor04,borucki11}. Driven by these discoveries, there is an increasing interest in the global properties of planetary systems, from the epoch of their formation in protoplanetary discs ({PPDs}) to their long-term stability. While there is a natural focus on {internal processes} that govern the {evolution} of such systems {\citep[e.g.][]{lee03,dullemond05,gorti09,morbidelli09,brasser09,blum10,williams11}}, it is clear that {not all planetary systems form in isolation and} environmental effects should be considered as well. Particularly, theoretical studies show that {external} photoevaporation {\citep{scally01,adams04, adams06, fatuzzo08}} and dynamical interactions {\citep{bonnell01,pfalzner05,olczak06,olczak10,spurzem09,lestrade11,dukes12,parker12,bate12}} can lead to the truncation of {PPDs} and planetary systems.

{While the external photoevaporation of PPDs has been studied observationally \citep{odell93,robberto08,rigliaco09,mann10}, there is no conclusive evidence of dynamical effects \citep{eisner06,olczak08,reche09}.} In part, this {is likely} due to the relatively short lifetimes of {PPDs} \citep[up to $\sim8$~Myr, e.g.][]{haisch01,hernandez08,ercolano11,smith12} compared to the time it takes stellar encounters to have an observable effect on the disc {population \citep[$\sim0.1$--$1$~Gyr, e.g.][]{adams10}. Previous observational studies on the Orion Nebula Cluster {(ONC)} \citep[e.g.][]{eisner06,olczak08} did aim to find traces of dynamical interactions in the population of PPDs, but lacked a sufficient number of sources and/or suffered from uncertainties on the disc mass measurements.}

We address the problem statistically by considering the sizes of {PPDs} as a function of {their} ambient stellar density, using samples of {PPDs} and young stellar objects {(YSOs)} from the {latest infrared surveys}. If stellar encounters truncate {PPDs} by tidally stripping the outskirts of the discs {\citep[e.g.][]{clarke93,heller95,hall96}}, {this should be observable} above some characteristic {ambient stellar} density {\citep[$\sim10^3~{\rm pc}^{-3}$][]{adams10}}, because the {encounter rate} increases with density {\citep[e.g.][Eq.~7-61]{binney87}}. In this Letter, we find model-independent evidence of {a change in the {PPD} size distribution for ambient stellar surface densities $\Sigma\simgreat 10^{3.5}~{\rm pc}^{-2}$} at the $>97\%$ confidence level.

%{Previous observational studies on the Orion Nebula Cluster \citep[e.g][]{eisner06,olczak08,mann10} drew very different conclusions ranging from not finding evidence for truncation of discs \citep{eisner06} to finding a correlation between disk size and proximity to the massive stars in the cluster in support of the photo-evaporation mechanism \citep{mann10}. \citet{olczak08} also investigated the correlation between the lack of infrared emission from individual stars, implying lack of circumstellar material, and the velocities of the sources pointing to dynamical interactions as an important mechanism for disc truncation. The main limitations of these studies are the low number of sources taken into account and the uncertainties related to the mass measurements.}

%We find model-independent evidence for {a change in the {PPD} size distribution for ambient densities $\Sigma\simgreat 10^{3.5}~{\rm pc}^{-2}$} {at the} $>2\sigma$ level. When compared to a simple model for the {dynamical truncation of {PPD}s} as a function of ambient density and age, the observed {change} appears in the {theoretically expected density range}. We then use the model to predict the maximum lifetimes of the habitable zone (HZ) in planetary systems as a function of their age and ambient density, by calculating the effect of dynamical interactions over much longer timescales than the typical {PPD} lifetime. We also provide a prediction for the {PPD} mass function at different ages and ambient densities.

\section{{Protoplanetary} discs and their environment} \label{sec:method}

\subsection{Data selection} \label{sec:data}
To verify whether a relationship exists between the sizes of {PPD}s and their {ambient stellar surface density, $\Sigma$, we} combine existing catalogues of {PPD}s and young stellar objects {in star-forming regions (SFRs) of} the solar neighbourhood. 

The data for the {PPD}s is taken from {circumstellardisks.org} {(Karl Stapelfeldt, NASA/JPL)}. This catalogue gathers resolved {PPD}s that have been confirmed and described in the literature. {If a {PPD} is resolved} in different wavelengths {\citep[probing different parts of the disc, see e.g.][]{lada06}}, the catalogue lists the largest {measured} diameter, implying that the disc radii used in this work are lower limits. {About $75\%$ of the {PPD} radii in our final sample {(see below)} are measured at wavelengths around $1~\mu{\rm m}$, with only $\sim25\%$ of the {PPD}s (all in low-density regions) being observed at mm wavelengths {(see Appendix~\ref{sec:app0})}. The {catalogue contains} an estimate for how well-resolved each disc is by listing the number of diffraction-limited beams that fit within its diameter. We only consider the discs for which this value is greater than unity. {This provides us with $133$ {PPD} sources from which we exclude those {whose host star is} classified as a main sequence star, weak-line T-Tauri star, {or} Class~0 YSO. Sources at distances $>500$~pc {(which covers all our YSOs) are also excluded}. These criteria reduce the sample to a total of $101$ sources.

To estimate the local ambient surface density of each {PPD} source, we use publicly available near-infrared data of nearby SFRs (see Table~\ref{tab:regions} in Appendix~\ref{sec:app0}). The ambient surface density of stars around each {PPD} is estimated as in \citet{casertano85} {-- see Appendix~\ref{sec:app0} for a detailed explanation}. The thus-obtained angular {ambient} surface densities are converted to {physical ambient stellar} surface densities, $\Sigma$, using the distances listed in Table~\ref{tab:regions}. In cases where the listed {PPD} distance differs from the distance to the region that it is a member of, we adjust the {PPD} distance and size. Since a low surface density can be due to incomplete YSO coverage, any discs with $\Sigma<0.1~{\rm pc}^{-2}$ are omitted from our analysis. {Moreover, the minimum {PPD} radius that can be resolved increases with distance and hence introduces a distance-dependent selection bias. To avoid this, we exclude any {PPD}s smaller than the smallest radius ($\sim50$~AU) that is resolved in the most distant region of our sample, which is the Orion Nebula Cluster (ONC). The final sample thus contains $67$ sources (see Table~\ref{tab:sources} in Appendix~\ref{sec:app0})}.}{ Completeness does not affect the densities because the surveys of Table~\ref{tab:regions} are complete down to the hydrogen-burning limit.}

\subsection{A simple theoretical estimate for the truncation of discs} \label{sec:theory}
To interpret the data, we include a rough theoretical estimate for the expected truncation radii of {PPD}s as a function {of $\Sigma$. This is obtained by combining the truncation induced by each particular encounter with the stellar encounter rate}. The derivation is presented in detail in {Appendix~\ref{sec:app1}.

We use the numerical simulations of disc perturbations in clustered environments by \citet{olczak06} to obtain the disc radius as a function of the encounter parameters. {We convert their expressions for the disc mass loss to a radial truncation assuming that it occurs by stripping the outer disc layers to the Lagrangian point between both stars, and adopting a PPD surface density profile $\Sigma_{\rm d}\propto r^{-1}$ \citep{olczak06}}. Under these assumptions, {we write for the upper limit to the disc radius}
\begin{equation}
\label{eq:rd}
r_{\rm d}(r_{\rm p},m_1,m_2)=\frac{r_{\rm p}}{\sqrt{m_2/m_1}+1} ,
\end{equation}
where $r_{\rm p}$ is the pericentre radius at which the perturber passes, $m_1$ the mass of the perturbed system, and $m_2$ the mass of the {perturber.} This approximation is validated in Appendix~\ref{sec:app1}.

The encounter radius $r_{\rm p}$ is obtained from the impact parameter $b$, encounter velocity $v$, and masses $m_1$ and $m_2$ by accounting for gravitational focussing {(see Appendix~\ref{sec:app1})}. The masses are assumed to follow a \citet{salpeter55} type initial mass function in the range 0.1--100~\msun, and the velocity distribution is taken to be Maxwellian {with a} velocity dispersion of $\sigma=2$~km~s$^{-1}$, as is typical for SFRs \citep{hillenbrand98,covey06}. This enables the derivation of the encounter rate as a function of $b$ and $v$ \citep{binney87}, which for a given age provides the total number of encounters $n$. Because encounters with pericentres at inclination angles $\theta\simgreat45^\circ$ with respect to the disc plane affect the disc only mildly, about $30\%$ of the $n$ encounters lead to the disc truncation described by Eq.~\ref{eq:rd} \citep{pfalzner05}. We use the probability distribution functions (PDFs) for $b$, $v$, and $m_2$ to calculate the PDF of the `most disruptive' encounter, i.e.~the parameter set that gives the smallest disc size, according {to \citet[Eq.~A5]{maschberger08}. We} then integrate the product of the disc radius $r_{\rm d}$, the PDF of the `most disruptive encounter', and the mass PDF of the perturbed object $m_1$ to obtain the expected disc truncation radius $r_{\rm tr}$. {To} compare the theoretical estimate to the observations, we relate the stellar volume density $\nu$ to the surface density $\Sigma$ as $\nu=\Sigma/2R$, where $R\approx 2$~pc is a typical radius for the SFRs in our sample \citep{hillenbrand98,evans09}. 

\section{Results}\label{sec:result}
\subsection{Evidence for environmental effects} \label{sec:theplot}
\begin{figure}
\center\resizebox{9cm}{!}{\includegraphics{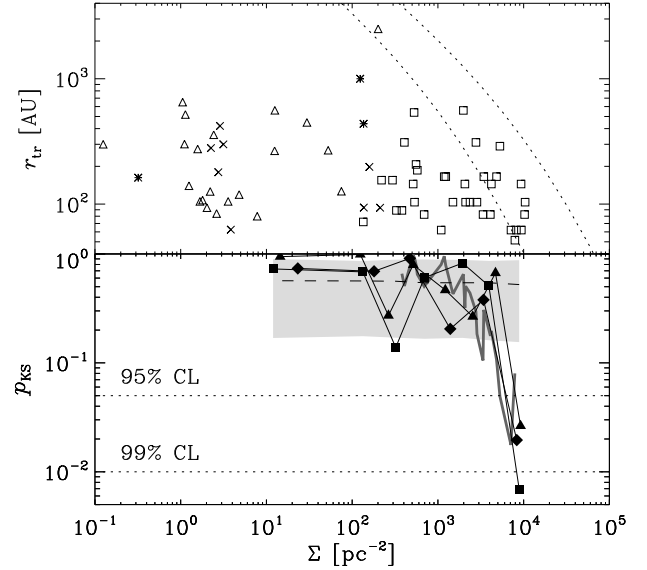}}\\
\caption[]{\label{fig:sizes}
      {{\it Top panel}: Protoplanetary disc (PPD) radius distribution versus ambient stellar density. Squares, triangles, stars, and crosses correspond to ONC sources, T-Tauri stars, Herbig stars, and YSOs, respectively. Dotted lines represent the theoretical truncation for ages of $0.3$ (top) and $1$~Myr (bottom). {\it Bottom panel}: KS test $p_{\rm KS}$-values for the PPD radii at that density to be consistent with the sample at lower densities, using $\{6,7,8\}$ (triangles, squares, diamonds) objects per density bin (see text). The grey thick line shows the $p_{\rm KS}$-value for ONC PPDs only when dividing the subsample in two at the indicated density. The horizontal dashed line and grey area indicate the median $p_{\rm KS}$ and its $1\sigma$ dispersion retrieved from 30,000 Monte Carlo experiments to verify the significance of the results (see text). Dotted lines mark $95\%$ (top) and $99\%$ (bottom) confidence levels.
                 }}
\end{figure}
The upper panel of Fig.~\ref{fig:sizes} {shows the observed {PPD}} radii  {versus $\Sigma$. The} distribution is} relatively insensitive to the ambient density {until $\Sigma\sim10^{3.5}~{\rm pc}^{-2}$, where it} {appears truncated at large radii. This is consistent with the simple theoretical approximation from Sect.~\ref{sec:theory}, which predicts a truncation at these densities for ages between $0.3$ and $1$ Myr} ({see} Fig.~\ref{fig:sizes}). The affected {PPD}s are all {in the ONC. If the truncation} is interpreted as being due to dynamical interactions, the theoretical curves {suggest that youngest sources in the ONC have ages} of $\sim0.6$~Myr, which is reasonably consistent with observations \citep{palla99,dario10,jeffries11}. {However, this is not a unique explanation, since the truncation might also be due to external photoevaporation {by nearby massive stars} \citep[e.g.][]{clarke07}. It is also important to keep in mind that some fraction of the {PPD}s at this projected surface density will actually reside in a region of lower volume density, either behind or in front of the high-density core of the ONC.} The distribution also shows some evidence of a density-independent {upper limit to the {PPD} radius} of $\sim10^3$~AU, which could be related to binarity \citep{artymowicz94,kraus12} or {be intrinsic} \citep{basu98}.

To test the statistical significance of the {change {in} the radius distribution,} we perform a Kolmogrov-Smirnov (KS) test, {which }is theory-independent {and} thus insensitive to model assumptions. {It gives the {probability $p_{\rm KS}$ that two samples were drawn from the same parent distribution}. Starting at the high-density end, we first divide the sample in density bins {of} $\{6,7,8\}$ objects per bin, which {represents a balance between good statistics} and enough bins to resolve the regime where the radius distribution changes}. The KS test is then carried out {comparing radii} in a bin at density $\Sigma_{\rm bin}$ with {those at lower densities (i.e.~$10^{-1}~{\rm pc}^{-2}<\Sigma<\Sigma_{\rm bin}$)}. We do not include bins at $\Sigma_{\rm bin}<10^1~{\rm pc}^{-2}$ to avoid low-number statistics in the reference sample. {The bottom panel of Fig.~\ref{fig:sizes} shows the results of the test}. At intermediate densities, {all} KS tests give high {$p_{\rm KS}$-values, but for densities higher than $\Sigma\simgreat10^{3.5}~{\rm pc}^{-2}$ there is} a pronounced drop. As such, the KS test yields {a detection} of a change in the {PPD} radius distribution {at the $>97\%$ confidence level}. {Note that this does not change when excluding the largest {PPD} in the sample (at $\Sigma\sim200~{\rm pc}^{-2}$). {The result} also holds within the ONC only {(grey} line in Fig.~\ref{fig:sizes}), when} dividing the ONC subsample in two at each density and running a KS test for the radii at both sides of the separation. {To check the result, we performed 30,000 Monte Carlo experiments in which the KS test was applied in the same way to distributions of randomly paired radii and surface densities, i.e. erasing any possible correlation. Figure~\ref{fig:sizes} indicates the resulting median $p_{\rm KS}$ and its $1\sigma$ dispersion, showing that the results obtained for the original sample are unlikely to be due to the adopted statistical method.}

\subsection{Implications for habitable zone occupancy lifetimes}\label{sec:hz}
\begin{figure}
\center\resizebox{8.5cm}{!}{\includegraphics{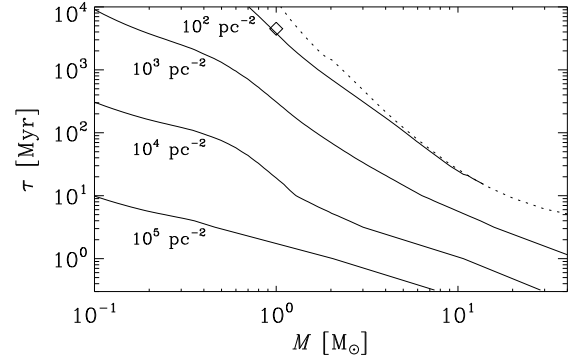}}\\
\caption[]{\label{fig:hz}
      {Habitable zone lifetime} as a function of stellar mass {for} different ambient stellar densities. The dotted line indicates the total lifetime of the host star, and the diamond indicates Earth. 
%{Note that a radius of 2~pc is used to convert volume to surface densities (see Sect.~\ref{sec:theory}).}                 
}
\end{figure}
If the drop of the $p_{\rm KS}$-value at high densities is indeed caused by dynamical truncation, then the model can be used to give the maximum {time} during which the {habitable zone (HZ)} can host a planet or a {PPD} {(`HZ lifetime')}. {We calculate} $r_{\rm tr}$ as a function of $\Sigma$ and $\tau$ as in Fig.~\ref{fig:sizes}, but without averaging over the host stellar mass to retain the mass dependence. We then compare it to the inner radius of the HZ, $r_{\rm HZ}^{\rm in}$, and determine until which age they overlap, {as a function of mass and ambient density}. The radius $r_{\rm HZ}^{\rm in}$ {depends on} the required radiative equilibrium temperature $T_{\rm eq}$, on the {stellar} luminosity $L$ and on the {properties} of the planet \citep[i.e.~$f$, a proxy for atmospheric thermal circulation, and the Bond albedo $A$, see][]{kasting93,tarter07}:
\begin{equation}
\label{eq:rhzmin}
r_{\rm{HZ}}^{\rm in}=\frac{1}{2} \sqrt{\frac{L f (1-A)}{4 \pi \sigma_{\rm{T}} T_{\rm{eq}}^4}} ,
\end{equation}
where $\sigma_{\rm{T}}$ is the Stefan-Boltzman constant. {Following \cite{borucki11}, we calculate $r_{\rm HZ}^{\rm in}$ for an Earth-like planet using $f=1$, $A=0.3$, and $T_{\rm eq}=307~K$, which is the maximum temperature that allows for the presence of liquid water when accounting for the greenhouse effect.} The luminosity is taken from stellar evolution models at solar metallicity \citep{marigo08}. {The results are} shown in Figure~\ref{fig:hz} for ambient densities of $\Sigma=\{10^2,10^3,10^4,10^5\}~{\rm pc}^{-2}$. The HZ lifetime decreases with density and stellar mass, due to the enhanced encounter rate and the large $r_{\rm HZ}^{\rm in}$, respectively. {These estimates for the HZ lifetime hold both for {PPD}s and planetary systems, {since} dynamical interactions {would have comparable effects in both cases} {\citep[e.g.][]{olczak06,parker12}}.} 
%Likewise, our results are unaffected by a possible change of the {PPD} radius with age.REF?? or out??}

%These estimates for the HZ lifetime hold both for {PPD}s and planetary systems, {since} dynamical interactions {would have comparable effects in both cases}{\citep[e.g.][]{olczak06,parker12}}. This also means that planet migration {does not strongly affect the {HZ lifetime}, because migration {takes place} only when the {PPD} is still present \citep{ida04}, and the encounter rate peaks due to the high ambient density.} Any planet that migrates beyond $r_{\rm tr}$ therefore has a high probability of being removed from the system in a next encounter. {Likewise, our results are unaffected by a possible change of the {PPD} radius with age.}

Figure~\ref{fig:hz} shows that, based on the Earth's existence alone, the solar system cannot have formed in a dense ($\Sigma>10^3~{\rm pc}^{-2}$) {environment}, unless the ambient density decreased on a short ($\tau\simless 100$~Myr) timescale. Conversely, meteoritic evidence indicates that the young solar system must have endured nearby supernovae \citep[e.g.][]{cameron77}, which provides a lower limit to the product {$\Sigma\tau\simgreat10^{3.8}~{\rm Myr}~{\rm pc}^{-2}$} \citep[see the review by][]{adams10}. {A plausible scenario is thus} that the solar system formed in a massive {($\sim10^4~\msun$, $\Sigma\sim10^3~{\rm pc}^{-2}$)}, but unbound association, which dispersed on a short ($\tau\sim10$~Myr) timescale \citep[see e.g.][although they refer to such a system as a `cluster']{dukes12}. Our results seem to disagree with \citet{eisner06} who derive the disc fraction {in the ONC} and find no evidence of disc truncations. {However, their conclusion may result from low-number statistics, and PPD mass estimates are more uncertain than radius measurements. On the other hand, our results agree with the studies of \citet{bonnell01} and \citet{spurzem09}, and would explain} why no planets have been found in the globular clusters {47~Tuc and} NGC~6397 \citep{gilliland00,nascimbeni12}, where $\tau\Sigma\sim10^8~{\rm Myr}~{\rm pc}^{-2}$ {within the half-mass radius}. This implies such a high number of encounters that it is {improbable} that {\textit{any}} bound planets survived, {most of them likely to have escaped the cluster due to two-body relaxation \citep[e.g.][]{kruijssen09c}}.

{We} note that our theoretical estimates are conservative and provide upper limits to the disc sizes, because (1) we do not account for potentially higher ambient densities in the past \citep[e.g.][]{bastian08b}, (2) {we neglect the presence of massive stars at earlier ages of the SFRs}, (3) we only consider the most disruptive encounters and {ignore} the cumulative effect of weak perturbations. Figure~\ref{fig:hz} thus provides strict upper limits to the HZ lifetimes.

\section{Further observational avenues} \label{sec:outlook}
We {present evidence} for a change in the {PPD} radius distribution at ambient stellar densities of $\Sigma>10^{3.5}~{\rm pc}^{-2}$ {at the $>97\%$ confidence level}, in line with the expected range due to close encounters with other stars {on a $\sim1$~Myr timescale}. These densities are only reached in the  densest parts of the ONC, which {is not only consistent with the detection of reduced {PPD} masses in the centre of the region \citep{mann10}, but also with} {studies concluding} that encounters are not important in the ONC as a whole \citep[e.g.][]{scally01}. Our results demonstrate that {the} stellar environment {can be} an important factor in setting the habitability of planetary systems. {For instance, the existence of unbound, free-floating planets \citep[see e.g.][]{bihain09,sumi11,strigari12} is a natural outcome of our results}. However, a ubiquity of Earth-like planets in the HZ {of stars remains likely} because a large fraction {($\sim90\%$)} of stars forms in unbound associations {(see \citealt{kruijssen12d} for a recent review, and observational references therein)}, of which the density quickly decreases after their formation.

\begin{figure}
\center\resizebox{8.5cm}{!}{\includegraphics{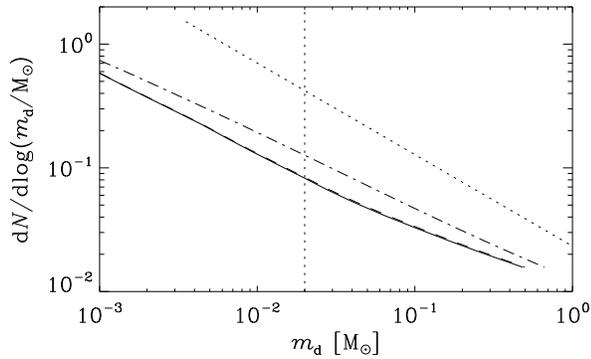}}\\
\caption[]{\label{fig:mdf}
      {{Evolution of the disc mass function (DMF) due to dynamical encounters at an age of 3 Myr for densities $\{2.25\times10^4, 2.5\times10^3, 5\times10^2\}~{\rm pc}^{-2}$ \textbf{(solid, dashed, dash-dotted)}, sampling NGC~3603 at \textbf{radius intervals of $\{0"-5",16.5"-43.5",46.5"-73.5"\}$. The dotted curve} shows the initial DMF, and the vertical dotted line indicates the $4\sigma$ sensitivity limit of ALMA Cycle~1 (30~min at 1.3~mm) at the distance of NGC~3603.}
                 }}
\end{figure}
To verify our results, more observations of {PPD}s in clustered environments are desirable. {A} fruitful approach would be to probe {PPD} truncations using disc mass measurements, {which would provide much larger samples of {PPDs. Figure~\ref{fig:mdf} shows the effect of (only) dynamical encounters on the disc mass function (DMF) using the disc mass loss description from \citet{olczak06} (see Appendix~\ref{sec:app2})}. {This} will be easily observable in dense and young stellar clusters with ALMA. While the full ALMA array will be able to directly measure disc sizes, we predict that, already from Early Science Cycle 1, the sensitivity will be sufficient to detect the variation of the DMF caused by the truncation (see Fig.~\ref{fig:mdf}), which would verify the result of Fig.~\ref{fig:sizes}.}

\begin{acknowledgements}
{We are grateful to the anonymous referee for a thoughtful and constructive report.} We thank Cathie Clarke, Michael Meyer, and Christoph Olczak for insightful comments on the manuscript, and David Jewitt, Thomas Maschberger, and Niels Oppermann for helpful discussions. NB is supported by the DFG cluster of excellence `Origin and Structure of the Universe' and HC by the Millennium Science Initiative, Chilean Ministry of Economy, Nucleus P10-022-F. 
\end{acknowledgements}

\bibliographystyle{aa}

\bibliography{mybib}

\appendix
\section{Properties of the sample} \label{sec:app0}
\begin{table}
\caption{Distribution of {PPD} sources over the host star type. `Unknown central star' indicates systems in which the central body has either not been detected or classified yet {(mostly proplyd silhouettes in the ONC)}. Young stellar objects are classified as such if their enhanced envelope emission suggests a younger age than T-Tauri or Herbig Ae/Be. {The `KS' column shows the sample after applying density and radius cuts for the KS test of Fig.~\ref{fig:sizes}.}}
\begin{center}
\begin{tabular}{ l | c | c}
\hline\hline
Type & Sources & KS\\ 
\hline
Herbig Ae or Be & 15 & 3 \\ 
T-Tauri & 39 & 21\\ 
Unknown central star (ONC) & 36 & 35\\ 
Young stellar object & 11 & 8\\ 
\hline
\end{tabular}
\end{center}
\label{tab:sources}
\end{table}

{After the selection procedure detailed in Sect.~\ref{sec:data}, we list the final {PPD} sample in Table~\ref{tab:sources}. {To assess the heterogeneity of the sample, we show the wavelengths of the radius measurements as a function of ambient stellar density in Fig.~\ref{fig:wavesigma}. The vast majority of sources ($75\%$) were measured in a narrow wavelength range below $3~\mu\rm{m}$ (i.e~near infrared wavelengths) with a spread of $0.25~\rm{dex}$ and centred at $1.32~\mu\rm{m}$. The remaining sources were measured at millimetre wavelengths. For the ONC sample, $80\%$ of all sources were measured at $0.66~\mu\rm{m}$ with very little scatter overall.} As shown in Fig.~\ref{fig:wavesigma}, the $16$ sources of the total sample measured at millimetre wavelengths all have ambient surface stellar density $<200~{\rm pc}^{-2}$. Therefore the distribution at densities above this value can be considered to be homogeneous.
\begin{figure}
\center\resizebox{8.5cm}{!}{\includegraphics{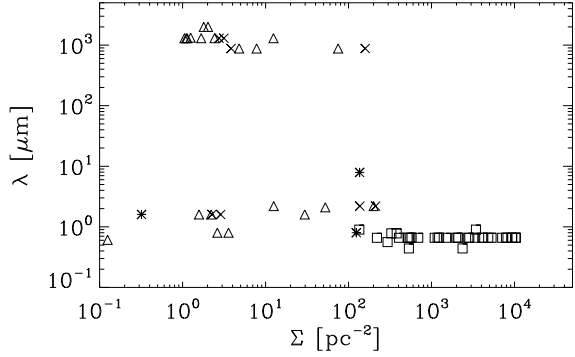}}\\
\caption[]{\label{fig:wavesigma}
      {Wavelengths $\lambda$ of the radius measurements for the final sample of protoplanetary discs versus ambient surface stellar density. Symbols have the same meaning as in the upper panel of Fig.~\ref{fig:sizes}.}}
\end{figure}

\begin{table}
\caption{Star-forming regions used in this study, listing the names of the regions, their numbers of objects $N_{\rm obj}$, and distances $D$. The fourth column shows the literature sources as (1) the Cores to Disks \textit{Spitzer} survey \citep{evans03}, (2) the Taurus \textit{Spitzer} survey \citep{rebull10}, (3) the \citet{robberto10} survey of the ONC.}
\begin{center}
\begin{tabular}{ l | c | c | c }
\hline\hline
Name & $N_{\rm obj}$ & $D$ (pc) & Survey \\ 
\hline
%Chamaeleon II & 29 & 180!! & 1 \\ 
%L673 & 27 & !! & 1\\ 
Lupus I & 20 & 150 & 1\\ 
Lupus III & 79 & 150 & 1 \\ 
Lupus IV & 12 & 150 & 1\\ 
Ophiuchus & 297 & 125 & 1\\ 
Orion Nebula Cluster & 7759 & 414 & 3 \\ 
Perseus & 387 & 250 & 1\\ 
Serpens & 262 & 415 & 1\\ 
Taurus & 249 & 137 & 2\\ 
\hline
\end{tabular}
\end{center}
\label{tab:regions}
\end{table}
The regions from which the YSOs are taken to estimate the ambient density are summarized in Table~\ref{tab:regions}. Using these samples, the ambient surface density of stars around each {PPD} is estimated as \citep{casertano85}:
\begin{equation}
\label{eq:surf}
\Sigma=\frac{N-1}{\pi d_N^2} ,
\end{equation}
where $N$ is the rank of the $N$th nearest neighbour, and $d_N$ is the projected angular distance to that neighbour. We use $N=20$, which is higher than the commonly-used value of $N=7$ \citep[cf.][]{bressert10} and is chosen to improve the statistics of the density estimates. {An additional effect of using a higher value of $N$ is a slight decrease of the density estimates. This should be kept in mind when comparing our densities those in other work.}

\section{A simple model for {PPD} truncations} \label{sec:app1}
In this Appendix, we derive the upper limit to the radii of protoplanetary discs ({PPD}s) due to dynamical encounters. Where appropriate, we emphasize that the derivation is conservative, such that the obtained truncation radius is indeed an upper limit.

\citet{olczak06} performed numerical simulations of disc perturbations and provided an expression for the relative disc mass loss $\Delta M/M$ due to encounters with other stars (their Eq.~4). If we assume that the mass loss occurs by stripping the outer disc layers and adopt the disc surface density profile {of $\Sigma_{\rm{d}}\propto r^{-1}$ used in their work, then} $\Delta r/r=\Delta M/M$. The expression for $\Delta r/r$ from \citet{olczak06} is consistent to within a factor of three with the scenario in which a disc is always truncated to the equipotential (Lagrangian) point between both stars. If the disc was already smaller than that radius, it is left relatively unperturbed. For the rough estimate made here, it thus suffices to write for the upper limit to the disc radius
\begin{equation}
\label{eq:rdapp}
r_{\rm d}(r_{\rm p},m_1,m_2)=\frac{r_{\rm p}}{\sqrt{m_2/m_1}+1} ,
\end{equation}
where $r_{\rm p}$ is the pericentre radius at which the perturber passes, $m_1$ is the mass of the perturbed system and $m_2$ is the mass of the perturber. The approximation of Eq.~\ref{eq:rdapp} {follows Eq.~4 of \citet{olczak06}} with reasonable accuracy for initial disc radii up to a few $10^3$~AU (consistent with the parameter space in Fig.~\ref{fig:sizes}), encounter distances $r_{\rm p}>0.002$~pc {(i.e. $r_{\rm p}/r_{\rm d}>0.2$)} and mass ratios $m_2/m_1>1$. We have verified that these conditions are satisfied for the encounters that are expected to determine the disc truncation (see below). Following \citet{binney87}, the impact parameter $b$ and the encounter radius due to gravitational focusing $r_{\rm p}$ are related as
\begin{equation}
\label{eq:focus}
b=r_{\rm p}\sqrt{1+\frac{2G(m_1+m_2)}{v^2r_{\rm p}}} ,
\end{equation}
where $v$ is the relative velocity of the encounter. This equation is inverted to derive $r_{\rm p}$ for each encounter.

The truncation radius $r_{\rm d}$ of Eq.~\ref{eq:rdapp} depends on the variable set $\{b,v,m_1,m_2\}$, for which we specify probability distribution functions (PDFs). For the masses, we use a \citet{salpeter55} type initial mass function in the range 0.1~\msun--$m_{\rm max}$, where $m_{\rm max}$ depends on age due to stellar evolution. For ages $\tau<4$~Myr we assume $m_{\rm max}=100~\msun$, while at later ages it is set by the \citet{marigo08} stellar evolution models at solar metallicity. The mass function is:
\begin{equation}
\label{eq:imf}
\Phi_m\propto\frac{\d N}{\d m}\propto m^{-2.35} ,
\end{equation}
which is normalized such that $\int\Phi_m\d m=1$. Assuming a Maxwellian velocity distribution, the total number of encounters per unit velocity $\d v$ and unit impact parameter $\d b$ {follows from the encounter rate ${\rm d}^2N/{\rm d}b{\rm d}v$ as} \citep{binney87}
\begin{equation}
\label{eq:number}
\Phi_{\cal N}\propto{\cal N}(b,v)\equiv\tau\frac{\d^2N}{\d b\d v}=\frac{8\pi^2\nu\tau b}{(4\pi\sigma^2)^{3/2}}\exp\left(-\frac{v^2}{4\sigma^2}\right)v^3 ,
\end{equation}
where $\nu$ is the local number density of stars, $\tau$ is the age of the region, and $\sigma$ is the velocity dispersion. The relative velocity ranges from $v=0$--$\infty$ and the impact parameter from $b=0$--$b_{\rm MAX}$ (see below). As in Eq.~\ref{eq:imf}, we have normalized $\Phi_{\cal N}$ such that $\int\!\!\!\int\Phi_{\cal N}\d b\d v=1$, by writing $\Phi_{\cal N}=f_{\rm dis}{\cal N}(b,v)/n$ and defining $n\equiv\int\!\!\!\int f_{\rm dis}{\cal N}(b,v)\d b\d v$ as the total number of encounters at age $\tau$. The factor $f_{\rm dis}\approx 0.3$ represents the fraction of encounters that leads to disc mass loss according to Eq.~\ref{eq:rdapp}. This accounts for the fact that encounters with pericentres at inclination angles $\theta>45^\circ$ with respect to the disc plane cause only weak mass loss and retrograde encounters leave the disc almost unperturbed \citep{pfalzner05}.

Given a sequence of encounters, the truncation of the {PPD} is set by the most disruptive encounter (\citealt{scally01}, although see \citealt{olczak06}), i.e.~$r_{\rm p,min}=f(b_{\rm min},v_{\rm min},m_{2,{\rm max}})$. If we assume that $\{b_{\rm min},v_{\rm min},m_{2,{\rm max}}\}$ are uncorrelated, implying that the region is not mass-segregated, the PDF of the most disruptive encounter becomes
\begin{equation}
\label{eq:pdis}
p(b_{\rm min},v_{\rm min},m_{2,{\rm max}})=p_b(b_{\rm min})p_v(v_{\rm min})p_m(m_{2,{\rm max}}) ,
\end{equation}
where $p_{\{b,v,m\}}$ represent the PDFs for the lowest $b$, lowest $v$ and highest $m_2$, respectively. Following the method of \citet[Eq.~A5]{maschberger08}, these three PDFs are defined as
\begin{eqnarray}
\label{eq:pbvm}
p_b(b_{\rm min})&=&n\Phi_b(b_{\rm min})\left(\int_{b_{\rm min}}^{\rm b_{\rm MAX}}\Phi_b(b\sp{\prime})\d b\sp{\prime}\right)^{n-1}\nonumber \\
p_v(v_{\rm min})&=&n\Phi_v(v_{\rm min})\left(\int_{v_{\rm min}}^{\infty}\Phi_v(v\sp{\prime})\d v\sp{\prime}\right)^{n-1}\\
p_m(m_{2,{\rm max}})&=&n\Phi_m(m_{2,{\rm max}})\left(\int_{m_{2,{\rm MIN}}}^{\rm m_{2,{\rm max}}}\Phi_m(m_2\sp{\prime})\d m_2\sp{\prime}\right)^{n-1}\nonumber 
\end{eqnarray}
where $\Phi_b(b)\Phi_v(v)\propto \Phi_{\cal N}(b,v)$ are the distribution functions for $b$ and $v$, {with $\Phi_b(b)\propto\nu\tau b$ and $\Phi_v(v)\propto\exp{(-v^2/4\sigma^2)}v^3/\sigma^3$}, again normalized to unity in both cases. In Eqs.~\ref{eq:pbvm}, $b_{\rm min}$, $v_{\rm min}$ and $m_{2,{\rm max}}$ indicate variable limits, and $b_{\rm MAX}$ and $m_{2,{\rm MIN}}$ indicate fixed limits. The fixed limit $b_{\rm MAX}$ represents the maximum impact parameter, which is given by the typical interstellar separation $b_{\rm MAX}=(48/\pi \nu)^{1/3}$ {\citep[the factor was chosen for consistency with][]{scally01}}. It should be noted that while this is a physically motivated choice, it only weakly influences the result since the {most likely most disruptive encounter} will typically be at $b_{\rm min}\ll b_{\rm MAX}$. Assuming an age of $\tau=1$~Myr, for surface densities of stars $\Sigma\leq 10^5$~pc$^{-2}$ we find that $p_b$ always peaks at impact parameters $b_{\rm min}>0.002$~pc {(i.e. $r_{\rm p}/r_{\rm d}\simgreat0.2$)}, whereas for $\Sigma\geq 10^{0}$~pc$^{-2}$ the {most likely most disruptive encounter} always has $m_2\geq0.5~\msun$, which after averaging over the mass function to account for the distribution of $m_1$ gives $m_2/m_1>2$. This validates the use of the approximation in Eq.~\ref{eq:rdapp}.

By combining the Eqs.~\ref{eq:imf} and~\ref{eq:pdis}, the total PDF is
\begin{equation}
\label{eq:phitot}
\Phi_{\rm tot}=p_b(b_{\rm min})p_v(v_{\rm min})p_m(m_{2,{\rm max}})\Phi_m(m_1) .
\end{equation}
It should be noted that we did not include the mass of the perturbed object $m_1$ in the PDF of the {most likely most disruptive encounter} (Eq.~\ref{eq:pdis}), but instead average over the mass PDF itself.  {The reason is that the stars in Fig.~\ref{fig:sizes} span a range of masses, and a `typical' relation between the truncation radius and ambient density is preferable}.

Combining the previous equations gives a theoretical estimate for the typical truncation radius $r_{\rm tr}$ as a function of the ambient density, velocity dispersion and age:
\begin{equation}
\label{eq:rtr}
r_{\rm tr}=\int\!\!\!\int\!\!\!\int\!\!\!\int_V r_{\rm d}\Phi_{\rm tot}\d b_{\rm min}\d v_{\rm min}\d m_1\d m_{2,{\rm max}} ,
\end{equation}
where $V$ indicates the complete phase space, i.e.~0.1~$\msun$--$m_{\rm max}$ in mass, 0--$\infty$ in velocity and 0--$b_{\rm MAX}$ in impact parameter. This expression provides the expected radius after the {`most likely most disruptive encounter'}, averaged over the stellar mass function to account for the unknown mass of the perturbed system.

\section{Evolution of the disc mass function} \label{sec:app2}
To calculate the evolution of the disc mass function (DMF), we assume that the initial disc mass $m_{\rm d,i}$ is related to the host stellar mass $m_1$ as
\begin{equation}
\label{eq:mdi}
m_{\rm d,i}=f_{\rm d}m_1 ,
\end{equation}
where $f_{\rm d}$ is a constant. We adopt $f_{\rm d}=0.03$, which is in good agreement with observations \citep{andrews05} and sufficiently accurate for the order-of-magnitude estimate made in Sect.~\ref{sec:outlook}. Using a \citet{salpeter55} stellar initial mass function (cf. Eq.~\ref{eq:imf}), the initial DMF is
\begin{equation}
\label{eq:nmdi}
\frac{\d N(m_{\rm d,i})}{\d m_{\rm d,i}}=\frac{1}{f_{\rm d}}\frac{\d N(m_1)}{\d m_1} ,
\end{equation}
{For each host stellar mass, we calculate the characteristics of the most likely most disruptive encounter as in Appendix~\ref{sec:app1}, using quantities that are appropriate for NGC~3603 (i.e.~$\tau=3$~Myr, $\sigma=4.5~{\rm km}~{\rm s}^{-1}$, and $R=1.45$~pc).} Given a certain encounter, the disc mass loss is calculated using the expression from \citet[Eq.~4]{olczak06}, which provides $\Delta\equiv\Delta m_{\rm d}/m_{\rm d}$ as a function of the host stellar mass $m_1$, the mass of the perturber $m_2$, the pericentre distance $r_{\rm p}$ and the disc radius $r_{\rm d}$. To account for the dependence of $\Delta$ on the radius, it is calculated for all radii from the observed sample at ambient densities $10^{-1}<\Sigma/{\rm pc}^{-2}<10^2$ (see Fig.~\ref{fig:sizes}), including those with $r_{\rm d}<50$~AU since the corresponding regions are all nearby and hence the detection limit is less stringent. At these densities the encounter rate is so low that the observed disc radii can be interpreted as `initial' radii. {The obtained values of $\Delta$ are then averaged to remove the dependence on $r_{\rm d}$, and integrated $\Phi_{\rm tot}$ (see Eq.~\ref{eq:phitot}) in the same way as $r_{\rm tr}$ in Eq.~\ref{eq:rtr}. This provides the expected relative mass loss as a function of host stellar mass $\langle\Delta\rangle$,\footnote{{Note that contrary to our Lagrangian approximation of Eq.~\ref{eq:rdapp}, the disc mass loss of the \citet{olczak06} equation does not increase monotonically with decreasing pericentre distance -- for very close encounters (typically $r_{\rm p}/r_{\rm d}<0.2$) the disc mass loss is reduced. In such cases, the most disruptive encounter is not the closest encounter, and we account for this by adjusting $r_{\rm p}$ to the value where $\langle\Delta\rangle$ peaks.}} and hence the final disc mass is approximately}
\begin{equation}
\label{eq:dmdfdmdi}
m_{\rm d}=\left(1-\langle\Delta\rangle\right) m_{\rm d,i} .
\end{equation}
The final DMF is then given by
\begin{eqnarray}
\label{eq:nmdf}
\nonumber \frac{\d N(m_{\rm d})}{\d m_{\rm d}}&=&\left(\frac{\d m_{\rm d}}{\d m_{\rm d,i}}\right)^{-1}\frac{\d N(m_{\rm d,i})}{\d m_{\rm d,i}}\\
&=&\frac{1}{\left(1-\langle\Delta\rangle\right)}\frac{\d N(m_{\rm d,i})}{\d m_{\rm d,i}} .
\end{eqnarray}

\end{document}